\documentclass[aps,prl,showpacs,twocolumn]{revtex4-1}
\usepackage{graphicx}
\usepackage[usenames]{color}
\usepackage{natbib}
\newcommand{\eq}[1]{Eq.~(\ref{#1})}
\newcommand{\fig}[1]{Fig.~\ref{#1}}
\newcommand{\be}[1]{\begin{equation}\label{#1}}
\newcommand{\ee}{\end{equation}}
\begin{document}

\title{Time-resolving intra-atomic two-electron collision dynamics }

\author{A. Emmanouilidou$^{1,2}$\email{}, A. Staudte$^{3}$, and P. B.
Corkum$^{3}$}

\address{
$^1$ Department of Physics and Astronomy, University College London, Gower Street, London WC1E 6BT, United Kingdom\\
$^2$Chemistry Department, University of Massachusetts at Amherst, Amherst, Massachusetts, 01003, U.S.A.\\
$^3$ Joint Laboratory for Attosecond Science, University of Ottawa and National Research Council, 100 Sussex Drive, Ottawa, Ontario, Canada K1A 0R6
}

\begin{abstract}
We generalize the one electron attosecond streaking camera to time-resolve the correlated two-electron escape dynamics during a collision process involving a deep core electron. The collision process is triggered by an XUV attosecond pulse and probed by a weak infrared field. The principle of our two-electron streak camera is 
that by placing the maximum of the vector potential of the probing field at the time of collision we get the maximum splitting of the inter-electronic angle of escape.  We thus identify the time of collision.

\end{abstract}
\date{\today}

\maketitle

 In conventional X-ray and collision physics the initial and final states of the fragments are the only accessible experimental observables.
 From differential measurements we have learned a great deal of what we know about the multi-electron structure of atoms. In these measurements, the correlated fragment dynamics is only accessed indirectly. 
 On the other hand, attosecond technology provides the tools to directly access correlated particle dynamics  \cite{Corkum2007Nature}. We show how to extend these attosecond measurement methods for resolving the intermediate stages of collisional processes.  
  
  Specifically, we generalize the attosecond streak camera \cite{ItataniPRL2002} to two escaping electrons. This allows us to  time-resolve the correlated two electron dynamics as the electrons leave the atom through the knockout mechanism \cite{TanisPRL1999} (sometimes called ``two-step-one") with the primary electron knocking out the secondary electron in a (e,2e) like process; the knockout process is observable as a result 
 of energetic electron or ion collisions or by absorption of an energetic XUV or X-ray photon.

 In the one electron streak camera \cite{Drescher2001Science, Uiberacker2007Nature,Eckle2008Science}, the electron's momentum is modified by the vector potential at the time the electron is ``born" into the streaking field. Measuring the distribution of the electron's final momentum we determine the range for  the electron's ``moment of birth". 
 One might expect that an analogous idea, applied to a variable measuring electronic correlation, would establish the range for the ``moment of intra-atomic collision". The angle between the momenta of the escaping electrons---inter-electronic angle, $\theta_{12}$---depends on both electrons and is thus a natural signature of electronic correlation.  $\theta_{12}$ is a variable of focus in collision physics and our variable of choice for formulating the principle of the two electron streak camera.

\begin{figure}[]
\centerline{\includegraphics[scale=0.5,clip=true]{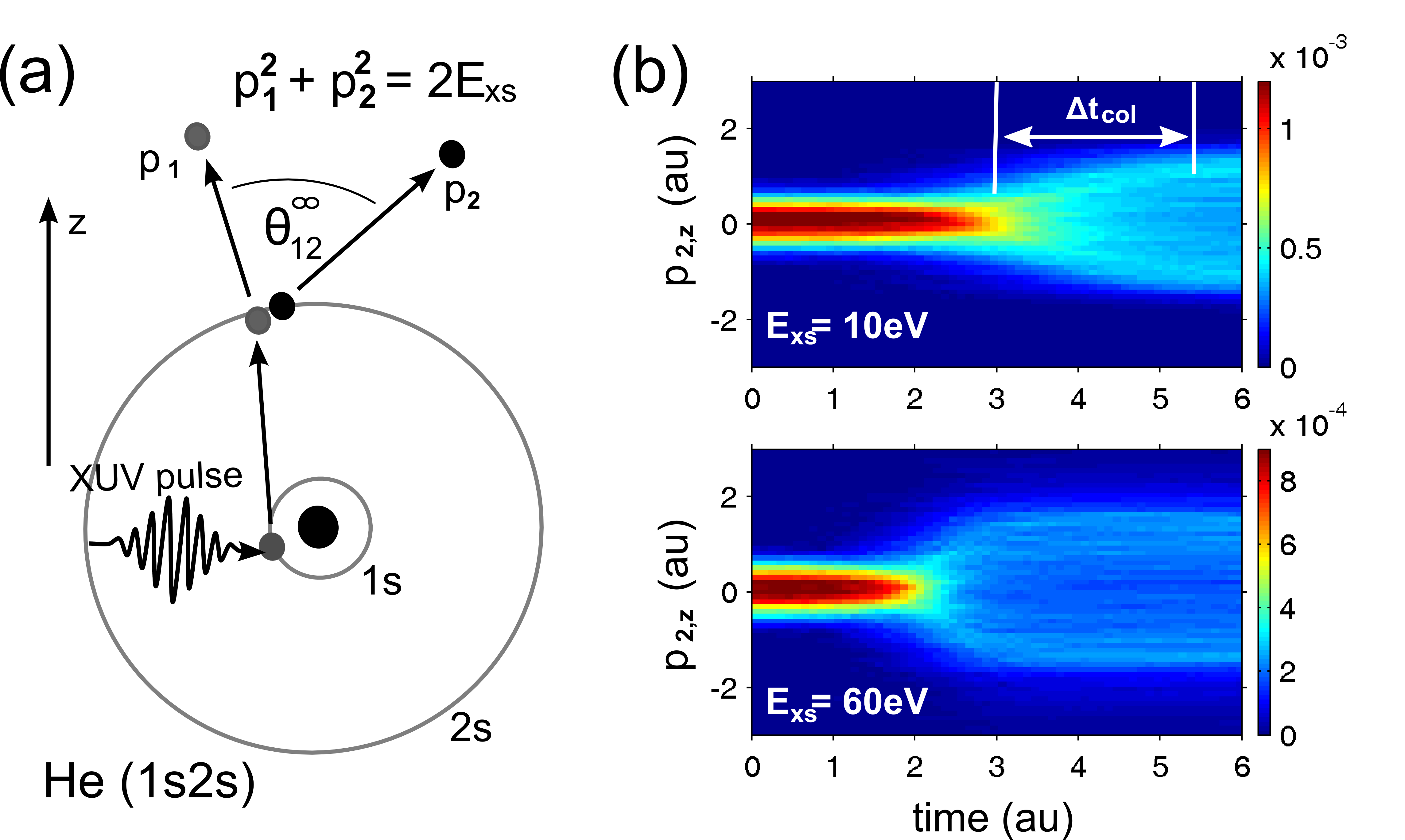}}
\caption{\label{fig:collisionmom}a) shows a schematic picture of the collision triggered by an XUV pulse at time $t=0$. b) shows the probability density for the z momentum component for the 2s electron for 10 eV (top row) and 60 eV (bottom row) excess energy. The sudden change in momentum $\Delta t_{col}$ takes place in the time interval between 3 and 5.5 a.u. for 10 eV and 2 and 3 a.u. for 60 eV.  }
\end{figure}

We demonstrate the concept of two-electron streaking using a He(1s2s) model system with the double ionization process triggered by an XUV-pulse. The 1s electron absorbs an XUV photon with energy above the double ionization threshold. This electron undergoes an intra-atomic collision with a 2s electron (\fig{fig:collisionmom} a) resulting in double ionization.
    In the absence of the streaking field the distribution of the inter-electronic angle of escape is peaked around a single angle. To probe the collision during the two electron escape we use a weak infrared optical field.
 When the streaking (probing) field is present, the final inter-electronic angle of escape changes as a function of the phase of the field.

  We find that $\theta_{12}$ exhibits a clear splitting around the most probable angle of escape in the absence of the streaking field. This splitting to lower and higher angles is due to the two different directions of the photo-electron with respect to the direction of the streaking field. (By photo-electron we mean the electron that initially absorbs the XUV photon.)
We measure the phase of the probing field with respect to the time the XUV attosecond pulse is applied. If we place the maximum of the vector potential at the time of collision we get the maximum splitting in 
$\theta_{12}$. This allows us to identify a characteristic time for the collision.

For our study  we use classical physics. Tracing classical trajectories in time allows us to clearly isolate the moment of collision. Classical models have been instrumental in understanding processes as diverse as non-sequential double ionization in strongly driven systems \cite{Corkum93PRL,Panfili2002PRL} and full fragmentation of three electron atoms triggered by single photon absorption \cite{EmmanouilidouPRA2007b,EmmanouilidouPRL2008}. For the latter process the three
 electrons escape through a sequence of attosecond momentum transferring collisions \cite{EmmanouilidouPRA2007b}. 
 
 Our model \cite{EmmanouilidouPRL2008} assumes that the 1s electron absorbs the photon at the nucleus \cite{ SchneiderPRL2002, EmmanouilidouPRA2007b, EmmanouilidouPRL2008}---an approximation that is exact in the high-energy limit. The initial conditions for the secondary electron (2s) are generated using the Wigner distribution \cite{HellerJCP1976} of the 2s hydrogenic orbital restricted on an energy shell. The energy shell corresponds to the ionization energy needed to remove the 2s electron from He(1s2s) \cite{EmmanouilidouPRA2007b}.
 We use a Classical Trajectory Monte Carlo phase space method \cite{AbrinesProcPhysSoc66a}. Regularized coordinates \cite{Kustaanheimo65JRAM} are used to avoid problems with electron trajectories starting at the nucleus. 
 
\fig{fig:collisionmom} a) also shows the release of the 2s electron from its bound state and the ultimate two electron escape taking place through an attosecond momentum transferring collision. This collision is clearly traced in the classical probability densities (see 
ref.\cite{EmmanouilidouPRA2007b})--the closest analog to a quantum mechanical density.    
 In \fig{fig:collisionmom} b) we show the classical probability density of the z momentum components for the 2s electron for excess photon energies, E$_{xs}$, of 10 eV/60 eV, where $E_{xs}=E_{\hbar \omega}-I$ and I is the fragmentation energy of He (1s2s). The sudden momentum change for the 2s electron 
 takes place approximately between times 3 and 5.5 a.u./2 and 3 a.u. for 10 eV/60 eV excess energy and  
 is a signature that a collision takes place. The transfer of momentum happens much faster with increasing excess energy resulting in an earlier time of collision. 
 We will show that the two-electron streak camera measures the time, $t_{col}$, that the sudden momentum change is complete. 
 
\begin{figure}[]
\centerline{\includegraphics[scale=0.6,clip=true]{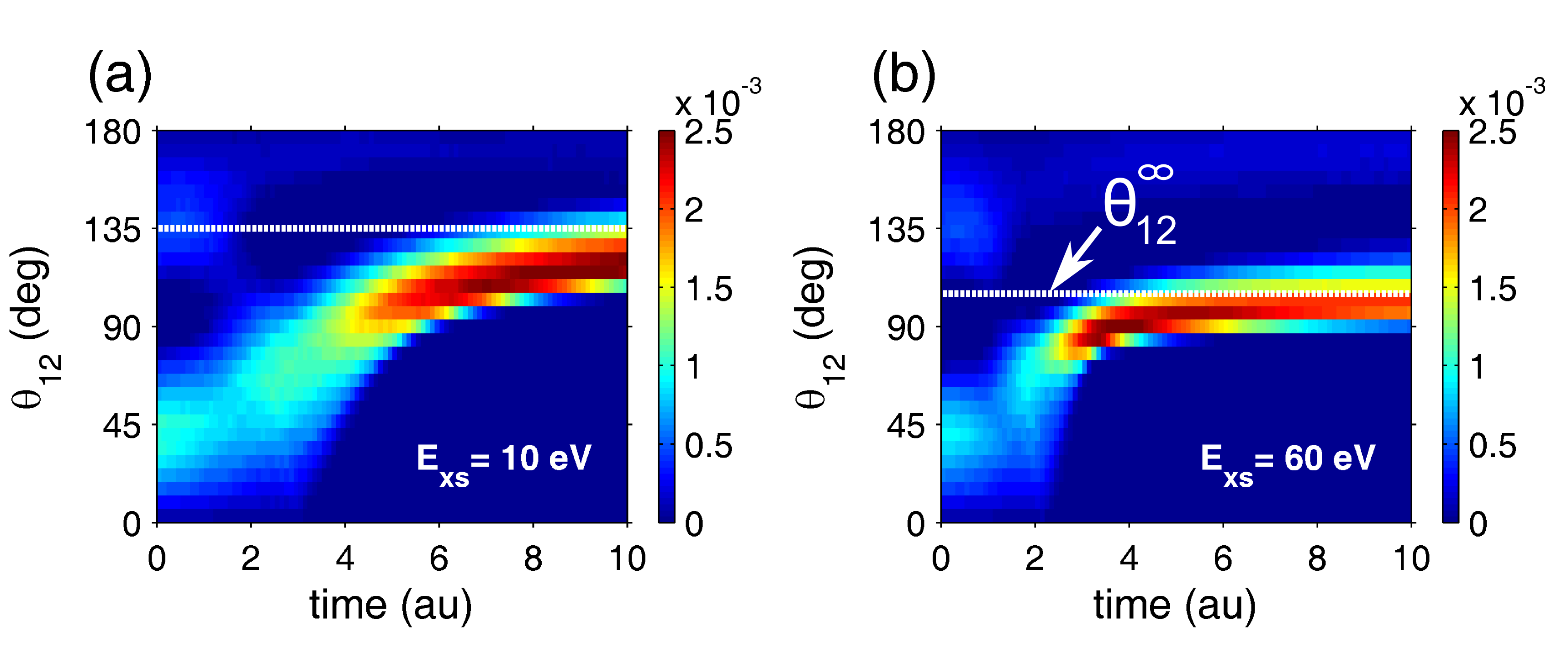}}
\caption{\label{fig:collisionangle}Probability density of $\theta_{12}$ as a function of time for photon-excess energy of 10 eV (a) and 60 eV (b). 
The sudden change in $\theta_{12}$ takes place in the time interval between 3 and 5.5 a.u. for 10 eV and 2 and 3 a.u. for 60 eV. $t_{col}\approx$5.5 a.u. /3 a.u. for 10 eV/60 eV. 
At very large times $\theta_{12}$ reaches the asymptotic value
of $\theta_{12}^{\infty}=135^{\circ}/105^{\circ}$ for 10 eV/60 eV, respectively.}
\end{figure}

 The collision between the two electrons is also visible in \fig{fig:collisionangle} when plotting the probability density for $\theta_{12}$ for 10 eV and 60 eV excess energies \cite{EmmanouilidouPRA2007b}. The sudden increase in $\theta_{12}$ as the two electrons move away from each other and escape from the nucleus is a signature of the two electron collision. For the higher excess energy the change in $\theta_{12}$ takes place much faster as is the case for the change in momentum (\fig{fig:collisionmom}). 
As shown in  \fig{fig:collisionangle}, the sudden change in $\theta_{12}$ takes place in the same time interval (3-5.5 a.u. for 10 eV and 2-3 a.u. for 60 eV) as the change in momentum shown in \fig{fig:collisionmom}. For very large times,  $\theta_{12}$ reaches the asymptotic value of $\theta_{12}^{\infty}=135^{\circ}/105^{\circ}$ for 10 eV/60 eV. Comparing \fig{fig:collisionmom} and \fig{fig:collisionangle} we see that $\theta_{12}$ is a much better observable for streaking than $p_{z}$ because $\theta_{12}$ has a much narrower asymptotic distribution.  
  
To probe the two-electron collision  we use a weak infrared laser pulse polarized along the z axis. The electric field is $E=E_{0}f(t)\cos(\omega t +\phi)$, where $\omega$ is the frequency, $\phi$ is the phase of the field and $f(t)$ is the pulse envelope. For our calculations we use $f(t)=1$ for $0<t<2T$ and $f(t)=cos^{2}((t-2T)\omega/8)$ for $2T<t<4T$. Time zero corresponds to the time the photon is absorbed from the 1s electron.  

Introducing the weak infrared laser field has two ramifications for the calculation. First, it requires solving a non-conservative driven three-body Coulomb system, in contrast to the single-photon process where the energy is conserved. Our propagation scheme has been generalized to account for laser-driven processes \cite{EmmanouilidouPRA2008a}. In addition, it breaks the spherical symmetry of the single-photo process. To account for the latter we slightly modify the initial phase space distribution along the lines of ref.\cite{SiedschlagJPB2005}---we take z to be the axis of polarization of the XUV pulse and weight the trajectories by a $cos^{2}\theta_{p_{1s}}$ dipole distribution for the photo-electron.

  We use a probing infrared pulse with frequency $\omega=0.0285$ a.u. and strength $E_{0}$=0.007 a.u./0.009 a.u. for 10 eV/60 eV excess energy. While $E_{0}$ and $\omega$ are not critical, we have chosen them so that the pulse does not significantly alter the attosecond collision and
 it does have an observable effect on $\theta_{12}$. The streaking field that we use corresponds to a very small ADK (Ammosov-Delone-Krainov) tunneling rate \cite{Delone91JOSAB} and so we can ignore tunneling of the 2s electron. Even if the tunneling were important, 
 it could be separated from the double ionization process by the low kinetic energy of the ATI (above threshold ionization) electrons just as is done in the single electron streak camera \cite{Drescher2001Science, Uiberacker2007Nature, Eckle2008Science}.

 Depending on the phase of the streaking field, the probe pulse can have a major effect on the asymptotic 
  \begin{figure}[h]
\centerline{\includegraphics[scale=0.56,clip=true]{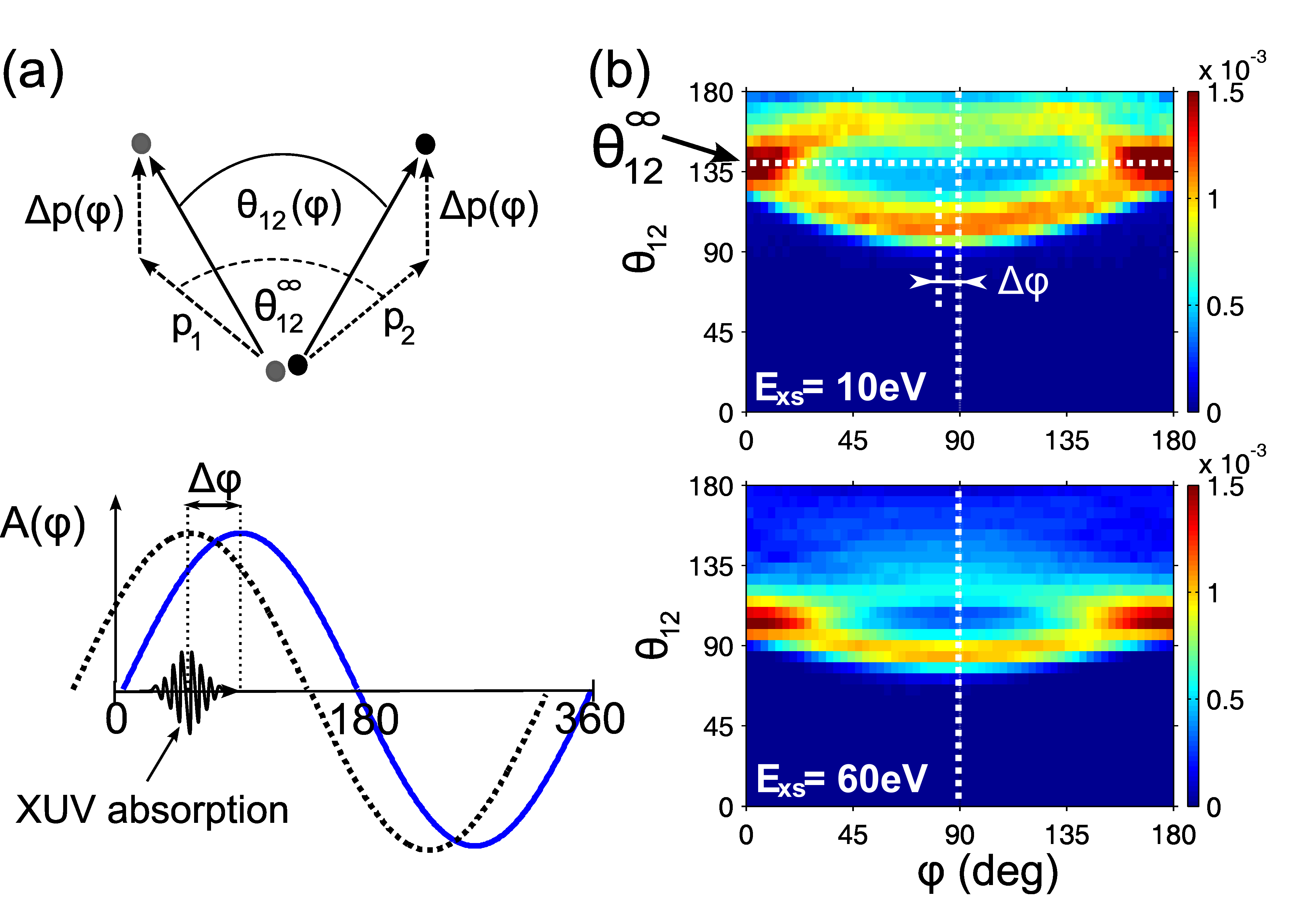}}
\caption{\label{fig:anglesprobepulse1}In a) we show that the streaking field causes a decrease in $\theta_{12}$ when the photo-electron is launched along the +$\hat{z}$ direction since adding $\Delta p$ to each of the electron momenta results in $\theta_{12}<\theta_{12}^{\infty}$. In b) we plot $\theta_{12}$ as a function of $\phi$ for 10 eV/60 eV (top/bottom row) excess energies in the presence of an XUV attosecond pulse and a weak infrared laser field. The parameters of the streaking field are $\omega=0.0285$ a.u. and $E_{0}$=0.007 a.u./0.009 a.u. for 10 eV/60 eV. $\Delta \phi$ is the shift of the maximum of the vector potential corresponding to a maximum of the split of $\theta_{12}$ as a function of $\phi$. $\Delta \phi=9^{\circ}/4.5^{\circ}$ for 10 eV/60 eV. }
\end{figure}

\noindent angular spectrum of the two escaping electrons.  As shown in \fig{fig:anglesprobepulse1} b), the probing pulse causes a clear splitting of the inter-electronic angle around $\theta_{12}^{\infty}=135^{\circ}/105^{\circ}$ (see \fig{fig:collisionangle}) for 10 eV/60 eV. The maximum split of $\theta_{12}$ occurs for a phase of the field $\phi_{delay}$ that is smaller than $\phi_{0}=90^{\circ}$. ($\phi_{0}$ is the phase corresponding to the maximum of the vector potential $A(\phi)$ at $t=0$, where  $-\frac{\partial \vec{A}(t)}{\partial t}=\vec{E}(t)$). Extracting the relevant information from \fig{fig:anglesprobepulse1} b) we find that $\Delta \phi =\phi_{0}-\phi_{delay}=9^{\circ}/4.5^{\circ}$ corresponding to $t_{col}=$5.5 a.u./2.76 a.u. for 10 eV/60 eV. The collision times we determine from \fig{fig:anglesprobepulse1} b)  agree with the collision times in \fig{fig:collisionmom} and \fig{fig:collisionangle}. 
  
 We now develop an analytical model to explain the split of the inter-electronic angle of escape as a function of the phase of the field and to show why $\Delta \phi$ gives the collision time. A small change in the photo-electron momentum has negligible impact on $\theta_{12}$. Therefore, we can neglect the interaction between the weak streaking field and the photo-electron before the collision. In addition, we assume that the transfer of energy from the 1s to the 2s electron is sudden. \fig{fig:collisionmom} and \fig{fig:collisionangle} show that this assumption is only approximately valid.   
The change in momentum for each electron due to the streaking pulse is given by:
 \begin{eqnarray}
 \Delta \vec p (\phi,t_{col})&=&-\int_{t_{0}}^{\infty}\vec{E}=-E_{0}\int_{t_{col}}^{\infty}f(t) cos(\omega t+\phi){\hat z} \nonumber \\
 &=&\frac{E_{0}}{\omega}\sin(\omega t_{col} +\phi)\hat{z}=-\vec {A} (\omega t_{col}+\phi).
\label{eq:1}
\end{eqnarray} 

  In the above we use the fact that the streaking pulse is zero at $t=\infty$. From \eq{eq:1} we see that the momentum change due to the streaking infrared field is along the positive $\hat{z}$ axis. $\Delta p$ depends on $t_{col}$
 and the phase, $\phi$, of the streaking field. Since $\Delta p>0$, the effect of the streaking field on the doubly ionizing events is different depending on the initial direction of launching of the photo-electron. As shown in \fig{fig:anglesprobepulse1} a),
 if the photo-electron is launched along the positive $\hat{z}$-axis then adding $\Delta p$ to each of the electron momenta results in $\theta_{12}<\theta_{12}^{\infty}$ thus accounting for the lower split of $\theta_{12}$ in \fig{fig:anglesprobepulse1} b). Similarly, if the photo-electron is launched along the negative $\hat{z}$-axis then subtracting $\Delta p$ from each of the electron momenta results in $\theta_{12}>\theta_{12}^{\infty}$ thus accounting for the upper split of $\theta_{12}$ in \fig{fig:anglesprobepulse1} b). From \eq{eq:1} we also see that the maximum split in $\theta_{12}$ occurs at $\omega t_{col}+\phi_{delay}=90^{\circ}=\phi_{0}$ resulting in $t_{col}=\Delta \phi/\omega$.
  
Our analytical model gives a very good estimate of the phase dependance of $\theta_{12}$ due to the presence of the streaking field, isolating the physics that underlies two-electron streaking. However, there remains another important issue. 
Classical physics has allowed us to select from all the trajectories that doubly ionize in the presence of the XUV and the streaking pulse only the trajectories that doubly ionize even in the absence of the streaking field (labeled below as ``doubles"). This selection
is not possible experimentally. We now discuss to what extent our two-electron streak camera is experimentally feasible.

 \begin{figure}[]
\centerline{\includegraphics[scale=0.6,clip=true]{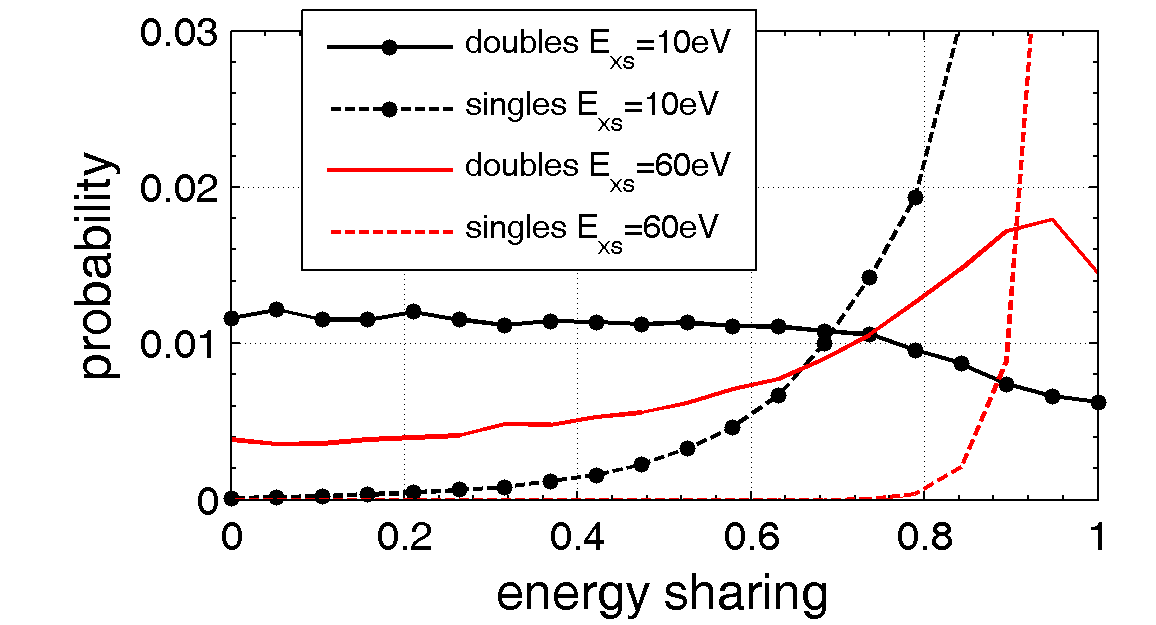}}
\caption{\label{fig:energysharing}The double ionization probability as a function of energy sharing averaged over all phases $\phi$ when the streaking field is on separately for the ``double" and `single" events for 10 eV/60 eV excess energy. The area under the ``double" events is 0.011/0.0084 while the area under the ``single" events is 0.013/0.0106. The sum of the two contributions, 0.024/0.019 for 10 eV/60 eV is the total double ionization probability in the presence of the XUV attosecond pulse and the streaking field.   }
\end{figure}
A small fraction of the events that singly ionize due to the XUV pulse involves the photo-electron exciting the 2s electron to Rydberg states. When the streaking field is subsequently turned on, it causes a fraction of the Rydberg states to ionize.  We label the events that doubly ionize due to the XUV
pulse alone as ``double" and the events that singly ionize due to the XUV pulse and subsequently doubly ionize due to the streaking field as ``single". The double ionization probability due to the XUV pulse is 0.011/0.0084 (``double" events). When the streaking field is turned on the double ionization probability increases to 0.024/0.019 with 0.011/0.0084 contribution from the ``double" and 0.013/0.0106 from the ``single" events for 10 eV/60 eV. These numbers were obtained by averaging over all phases $\phi$. There is a phase dependance of the double ionization probability of ``singles" and of their momentum distribution. While not the focus of this paper, the ``singles" provide a measurable that can be exploited
to characterize the bound state electron wave-packet created by the internal two-electron collision.

``Single" trajectories can be separated from the ``doubles" if one considers the energy sharing, $\frac{|\epsilon_{1}-\epsilon_{2}|}{|\epsilon_{1}+\epsilon_{2}|}$, with $\epsilon_{1}$, $\epsilon_{2}$ the asymptotic energies of the 1s and 2s electrons. \fig{fig:energysharing} shows that the ``single" trajectories typically have a high asymmetry in their energy sharing with $\frac{|\epsilon_{1}-\epsilon_{2}|}{|\epsilon_{1}+\epsilon_{2}|}>0.85/0.95$ for 10 eV/60 eV excess energy. In contrast, ``doubles" have a much smaller asymmetry. By considering only trajectories with asymmetry in energy sharing less than 0.85/0.95 in \fig{fig:energysharing} for 10 eV/60 eV excess energy, we separate 64/70 \% of the ``singles" while losing only 9\% of the ``double" events we want to probe. If, in addition, we use a cut-off in the inter-electronic angle, without further loss of ``double" events, we isolate 70/80\% of the ``single" events. Both of these procedures for isolating the ``single" events are available to an experimentalist.

Concluding,  regarding the ``double" events, so far we have exploited one variable of the collision, but a lot of unused information remains. A promising direction for beginning the process of unravelling more information about the collision 
is to consider how the inter-electronic angle of escape depends on the phase of the field for different energy sharing between the two electrons.
  We expect that, just as the single electron streak camera can be generalized to FROG (Frequency Resolved Optical Gating) \cite{MairessePRA2005}, the two-electron streak camera can also be generalized to give us a full picture of the intra-atomic collision.

 Finally, as attosecond free electron lasers (FEL) develop, the two-electron streak camera can be extended to deep core electron dynamics. 
We also expect that the streak camera concept can be applied to many electron intra-atomic \cite{EmmanouilidouPRA2007b} and intra-molecular collisions, one of the exciting problems at the frontier of attosecond science. 

The author A.E gratefully acknowledges funding from EPSRC under Grant No. EPSRC/H0031771, from NSF, Grant No. NSF/0855403 and Jan Michael Rost and the Max Planck Institute for Complex Systems
for access to their computational resources.

\bibliography{BibAPAPRL20101.bib}
\bibliographystyle{unsrt}

\end{document}